# Colloidal stabilization of graphene sheets by ionizable amphiphilic block copolymers in various media

*Maria-Teodora Popescu [a], Dimitrios Tasis [a,b], Konstantinia D. Papadimitriou [a], Sandra Gkermpoura [a,c], Costas Galiotis [a,c] and Constantinos Tsitsilianis [a,c]\**

[a] Foundation of Research and Technology Hellas, Institute of Chemical Engineering Sciences (ICE-HT), P.O. Box 1414, 26504 Rio Patras, Greece

[b] Department of Chemistry, University of Ioannina, 45110 Ioannina, Greece

[c] Department of Chemical Engineering, University of Patras, 26504 Rio Patras, Greece

\*Corresponding author. E-mail: ct@chemeng.upatras.gr (Constantinos Tsitsilianis)
Tel: 0030 2610969531

**Abstract**

In this work, linear polystyrene-poly(2-vinylpyridine) (PS-b-P2VP) and heteroarm star $PS_{22}P2VP_{22}$ ionizable block copolymers were used as dispersing agents for the liquid-phase exfoliation of pristine graphene. Various strategies such as direct exfoliation, film hydration and phase transfer have been employed and compared with. The best strategy involved a two-step process, namely, pre-exfoliation of graphite in polymer/$CHCl_3$ solutions followed by phase transfer to acidified water. High concentrations of stable aqueous suspensions of graphene flakes, highly enriched in monolayer structures, were then obtained by using the star-shaped copolymers as stabilizers. The as-prepared graphene/copolymer hybrids were used as a filler material in order to prepare functional polymer composites for mechanical reinforcement. Such copolymer-modified graphene sheets have proven to be efficient reinforcing agents of PVA, as a significant increase of storage modulus (145% higher than that of neat PVA) was achieved even at a low graphene weight fraction of 0.1 wt%.

**1. Introduction**

Since the seminal work of monolayer graphene isolation through the "scotch-tape" method [1], this two-dimensional allotropic form of carbon has received a great deal of attention in a wide variety of research fields.[2] Its unique one-atom-thick structure composed of solely $sp^2$-hybridized carbon atoms is responsible for the extraordinary mechanical, electrical and optical properties, which, in turn, make graphene a potential candidate to replace conventional materials in a number of emerging applications. These include polymer composites [3], energy conversion and storage [4,5], catalysis [6], etc. For specific applications, stable suspensions of efficiently exfoliated sheets at high concentration are often needed. So far, different strategies have been developed for the preparation of graphene dispersion in either organic or aqueous media.[7-12] In particular, the most common approach involves the oxidative exfoliation of graphite flakes with subsequent production of the so-called graphene oxide (GO).[13] However, the harsh conditions required for the oxidation reaction are responsible for the generation of high density defect sites onto the basal plane of graphene, rendering the carbon nanostructure an insulator. The structural properties of the oxygenated form of graphene may be partially restored through either thermal or chemical reduction strategies.[14] The resulting chemically converted graphene is readily suspended in various media only in the presence of surfactant-like substances, but cannot be considered as structurally intact graphene. Due to the aforementioned disadvantages of the oxidation/reduction approach, alternative methods have been explored towards the production of defect-free graphene suspensions.

The most common strategy involves the sonication-assisted direct exfoliation of graphite flakes in various liquid media, which results in the preparation of stable suspensions highly enriched in few-layer graphenes (n < 5).[7,8,15-17] As has been demonstrated earlier, graphene can be efficiently dispersed in liquid media, in which their surface energy matches that of graphene, c.a. ~68 mJ/m$^2$ (or in terms of surface tension γ~40 mJ/m$^2$).[7,8] However, the solvents that fulfill this criterion are few, e.g. N-methyl-2-pyrrolidone (NMP), 1,2-dichlorobenzene (DCB) or dimethylformamide (DMF). More importantly, the aforementioned media exhibit high boiling points, an issue which restricts their applicability as exfoliation agents. In another approach, based on the minimization of the enthalpy of mixing, the graphene concentration ($C_G$) depends on the graphene/solvent Flory-Huggins interaction parameter ($\chi_{GS}$), with the trend being the lower the $\chi_{GS}$, the higher the $C_G$. In terms of Hildebrand solubility parameters, the graphene concentration may be maximized by matching the solubility parameters of graphene ($\delta_G \approx 21.5$ MPa$^{1/2}$) with that of the solvent, $\delta_S$, since it is well known that $\chi_{GS} \sim (\delta_G - \delta_S)^2$. The same correlation is valid also in terms of Hansen solubility parameters.[8,18] The advantage of the solvent-based exfoliation of graphite is that defect-free graphene platelets can be obtained. However, the mean size of the suspended graphenes is limited to sub-micrometer values, due to the utilization of extended sonication times [19] and the exfoliated material is usually a few-layer graphene, which could, in effect, yield nanocomposites of inferior mechanical properties as compared to those that incorporate single graphene layers.

In order to exfoliate and stabilize graphene in either low boiling point organic or aqueous media, which do not fulfill the solubility parameter criteria, homopolymers have been used as dispersing agents according to the steric stabilization concept.[20,21] Stabilization of the exfoliated graphene may occur when part of the polymer chain physically adsorbs onto the graphene nanosheet surface while the other part of the chain interacts with the solvent molecules. In the work of Coleman and co-workers [20], a simple phenomenological model was proposed which correlates the concentration of dispersed graphene with the interaction parameters between graphene-polymer, $\chi_{GP}$, and solvent-polymer, $\chi_{SP}$, both of which having to be minimized in order to maximize $C_G$. In terms of Hildebrand solubility parameters of components (graphene, $\delta_G$, polymer, $\delta_P$, and solvent, $\delta_S$), the stabilized graphene concentration will be maximized if the solubility parameters match, i.e. $\delta_G \approx \delta_P \approx \delta_S$.

Another strategy to stabilize graphene dispersions in liquid media involves the utilization of amphiphilic substances, including either low molecular weight surfactants, or macromolecules of the type of block copolymers.[7] Stabilization of graphene is thus relied on favourable interactions that take place between graphene surface and the solvophobic domain, and those between the solvent and the solvophilic part of the amphiphilic stabilizer. This results to physisorption of a specific domain of the amphiphile onto the nanosheet surface, while the solvophilic part protrudes into the solvent, forming micelle-like structures. In the case of macromolecular amphiphiles, such as block copolymers of A-*b*-B type, additional factors dealing with the macromolecular features, such as chemical composition, the molecular weights of the blocks and the macromolecular architecture should affect the stabilization mechanism. The existence of four components (graphene, solvent, A-block, B-block) increases the number of interaction parameters that have to be taken into account to six. In addition, the high diversity of the macromolecular characteristics, makes predictions of choosing the best copolymer stabilizer quite difficult. However, by choosing block copolymers with highly incompatible A/B blocks, that is A highly hydrophobic and B highly hydrophilic (polyelectrolyte type), the main interactions that govern the stabilization could be reduced to those between graphene(G)/A-block, $\chi_{GA}$, and solvent/B-block, $\chi_{SB}$, since the interactions between G-B, G-S, S-A and A-B are minimized. Thus, at a first approximation, it could be assumed that the solubility parameter pairs of graphene/A-block, and solvent/B-block should match, i.e. $\delta_G \approx \delta_A$ and $\delta_S \approx \delta_B$, in order to achieve reasonably high graphene concentrations. However, we should note that a block copolymer in a selective solvent (poor solvent for the one block) will self-associate, forming micelles which might be antagonistic to graphene exfoliation and stabilization.

Indeed, some past studies have shown that sonication-assisted exfoliation of pristine graphene sheets in the presence of block copolymers results in the formation of stable graphene suspensions in either aqueous,[22-25] polar organic [23,24] or low polarity media.[26] More recently, Tagmatarchis and co-workers [23] reported the use of a symmetric polystyrene-*b*-poly(2-vinyl pyridine) block copolymer (PS-*b*-P2VP, 56 wt% P2VP), as steric stabilizer of graphene. These authors managed to disperse the graphene/copolymer hybrid nanostructures in aqueous environment by pre-exfoliation of graphite flakes in the water-miscible solvent, NMP, followed by dilution with acidified water.

Herein, we demonstrate novel methodologies for acquiring pristine graphene suspensions in different liquid media, i.e. low boiling point organic solvents, aqueous environment or ionic liquid. High quality graphene nanosheets were obtained by using asymmetric ionogenic block copolymers (c.a. 80 wt% fraction of P2VP, a potentially hydrophilic block through protonation) of different macromolecular architectures i.e. linear PS-*b*-P2VP and $PS_n$-$P2VP_n$ heteroarm star copolymers (Figure 1).[27-29] It should be mentioned that amphiphilic star-shaped copolymers have been successfully utilized as "smart" dispersing agents of carbon nanotubes (MWCNT) in aqueous media.[30] To this end, three processing strategies have been developed and compared with, in order to assess the exfoliation efficiency of graphene suspensions, by using block copolymer stabilizers. Moreover, a systematic study was performed concerning the effect of various processing parameters governing the exfoliation efficiency of graphene in solution, such as, polymer concentration, solution acidity etc.

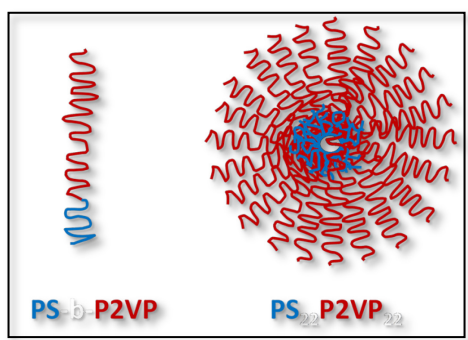

**Figure 1.** Schematic representation of the linear and the heteroarm star copolymers involved in this study.

More precisely, pre-exfoliated graphene sheets in chloroform were successfully phase-transferred into acidic aqueous media and subsequently to hydrophobic ionic liquids. Beside the direct exfoliation and the shuttle transfer process between immiscible media, film hydration was used as an additional protocol for preparing stable graphene suspensions. Graphene dispersability, as well as, exfoliation efficiency were assessed for all three processing strategies and for the different macromolecular topologies, by transmission electron microscopy (TEM) and Raman spectroscopy. The phase transfer route seems to be the most promising as it leads to graphene nanosheets, appreciably enriched with monolayer structures (preliminary results have been reported in a recent rapid communication[31]). Finally, the highest quality graphene/star polymer hybrids, suspended in aqueous media by phase transfer, were used to evaluate their ability as reinforcing agents in polymer-based nanocomposites by using poly(vinyl alcohol) (PVA), as a model polymeric matrix.

## 2. EXPERIMENTAL SECTION

### 2.1 Materials and reagents

Unless otherwise stated, reagents and solvents were obtained from Aldrich and were used as received. Graphite with an average particle size of 500 μm and a purity of >95% was supplied by NGS Naturgraphit GmbH (batch: large flakes). Both linear and heteroarm star block copolymers, comprising PS and P2VP blocks/arms, were synthesized by living anionic polymerization via sequential addition of monomers and under inert atmosphere (Ar slight overpressure) in tetrahydrofuran, in the presence of LiCl, according to standard procedures.[32] Briefly, for the PS-b-P2VP linear block copolymer, secondary butyl lithium (s-BuLi) was used as the initiator to polymerize styrene, followed by 2-vinyl pyridine (2VP) addition. As far as the heteroarm star copolymer is concerned, the so-called multi step "in-out method" was pursued. The first generation of PS arms were formed in the first step, by the same procedure followed for the linear counterpart. Subsequently, these "living" linear PS chains were used as macroinitiators for the polymerization of a small amount of divinylbenzene (DVB) acting as a crosslinker forming a tight polyDVB core. A living PS star-shaped polymer was thus formed, bearing within its polyDVB core an equal number of active sites with its PS arms. The PS star precursor was isolated and characterized. These sites were used to grow the second generation of P2VP arms upon the addition of 2VP, yielding the $PS_n$-$P2VP_n$ heteroarm star copolymer.[27] All samples have been characterized by a combination of gel permeation chromatography (GPC), $^1$H NMR, and light scattering and the molecular weight data are summarized in Table 1.

Table 1. Molecular characteristics of the PS-P2VP copolymers

| Polymer | Topology | No of arms | $M_{w, PS\ arm}$ gr/mol | $M_{w, P2VP\ arm}$ gr/mol | P2VP (mol%) |
|---|---|---|---|---|---|
| $PS_{26}$-b-$P2VP_{204}$ | linear | 1+1 | 2700 | 21 450 | 88.7 |
| $(PS_{35})_{22}$-$(P2VP_{136})_{22}$ | star | 22+22 | 3600 | 14 300 | 79.5 |

**2.2 Direct copolymer-assisted exfoliation of graphite**

Exfoliation of graphite in copolymer solutions was carried out in the following media: acidic water (pH 2), methanol, ethanol, isopropanol, isobutanol, chloroform and DMF. Graphene suspensions with two different polymer concentrations were prepared, that is 0.1 and 1.5 mg/mL. Note that these concentrations were chosen as to be equimolar for the linear and star copolymer, respectively. Pristine graphite flakes were added into the corresponding medium (starting graphite concentration 2 mg/ml) and the dispersion was subjected to sonication for three cycles of 30 min, using a tip sonicator (Q55 QSonica, 55 Watts, 20 kHz) at 10% amplitude. Note that the polymer remained intact after this treatment as proved by GPC (see ESI). During sonication, graphene sheets were exfoliated from the graphite flakes and dispersed homogeneously in the aqueous or organic medium. The suspensions were then left to settle overnight. In order to discard the graphite flakes, the upper phase (~80% of the total volume) was carefully pipetted away and was centrifuged at 2000 rpm for 30 min. From this suspension, the supernatant part was taken and used for further measurements.

## 2.3 Film hydration

Sonicated $CHCl_3$ dispersions of graphene/polymer mixtures were centrifuged for 30 min at 2000 rpm and the supernatant part was carefully pipetted and transferred to a different vial. Further on, the $CHCl_3$ phase was slowly evaporated by heating at 30 $^0$C overnight, followed by vacuum drying for 3 h. The obtained film was hydrated with an equal volume of water at pH 2. The aqueous dispersion was subjected again to sonication for 30 min followed by centrifugation at 2000 rpm for 30 min.

## 2.4 Two-phase shuttle transfer

On the top of a graphene/polymer centrifuged suspension prepared in $CHCl_3$, an equal volume of acidic water at pH 2 was added as the receiving compartment. After moderate agitation (100 rpm) for 72h, the aqueous phase was black-colored and was isolated for further measurements. In a subsequent step, the pH of the aqueous media was increased to ~ 6.8, upon addition of appropriate volume of NaOH 0.1M (~20 μL). Further agitation for 24h resulted in diffusion of the hybrids to the $CHCl_3$ phase. The total time of phase transfer was considered as the time when no further increase in the optical absorbance at 660 nm was detected. Graphene/polymer hybrid dispersed in acidic water was contacted with an equal volume of 1-butyl-3-methylimidazolium hexafluorophosphate [BMIM][PF$_6$] hydrophobic ionic liquid. After moderate agitation (60 rpm) for 3 min, the ionic liquid phase was black-coloured.

**2.5 Quantitative estimation of graphene/polymer hybrid concentration**

Estimation of graphene/polymer hybrid concentrations determined by UV–vis spectroscopy at a wavelength of 660 nm with a Shimadzu UV 2500 absorption spectrophotometer. For this purpose the apparent absorption coefficients of graphene/polymer hybrids ($\alpha_{G/P}$) in various solvents were determined by the formula $\alpha_{G/P} = A/l\ c_{G/P}$, which obeys the Lambert–Beer law (Figures S1 and S2). The results for various systems are presented in Tables S1 and S2.

**2.6 Raman Spectroscopy characterization**

MicroRaman (InVia Reflex, Renishaw, UK) spectra were recorded with 514.5 nm (2.41 eV) excitation (the laser power was kept below 1 mW). For the preparation of samples, diluted graphene suspensions (concentration ~ 0.01 mg/ml) were spin-coated onto Si/SiO$_2$ wafers.

**2.7 Electron microscopy**

TEM images were obtained using a JEM-2100 transmission electron microscope operating at 200 kV. A drop of a diluted suspension was placed on the top of a formvar-coated carbon grid (Agar Scientific). The solvent was gently absorbed away by a filter paper. The grids were then allowed to air dry at room temperature prior to observation. Field Emission Scanning Electron Microscope (FESEM) images were recorded with a ZEISS SUPRA 35VP device, whereas the deposited samples were sputtered with gold prior to observation. For the nanocomposites (see below), cryofractured samples were gold sputtered and then scanning electron micrograph (SEM) images were taken by using a LEO Supra 35VP microscope.

## 2.8 Dynamic Mechanical Analysis (DMA)

Dynamic mechanical analysis was carried out with a dynamic Q800-TA mechanical analyzer. The specimens were cut in the form of rectangular films (20x5x0.1 mm) and were deformed at a constant frequency of 10Hz in an inert nitrogen atmosphere. The temperature scans were carried out at a rate of 3 $^{o}$C/min within the temperature range of -50 – 170 $^{o}$C.

## 3. RESULTS AND DISCUSSION

### 3.1 Exfoliation of graphite: quantitative results

*3.1.1 Direct exfoliation*

The colloidal stability and exfoliation efficiency of carbon nanostructures in the presence of polymeric stabilizers were assessed firstly in a wide variety of media, both organic and aqueous (see Experimental Section). With regards to the chemical affinity of the solvents, the acidic aqueous and alcoholic media are considered selective for the P2VP segments, whereas chloroform and dimethylformamide are common solvents for both PS and P2VP blocks. As reference samples, graphene flakes were isolated in neat solvents under sonication. No appreciable exfoliation of graphene was observed under such conditions.

Regarding the solubility of both neat copolymers in acidic water, turbidity was observed, which was more pronounced at higher polymer concentration. This indicated the formation of relatively large aggregates due to self-assembly of the polymer chains in the aqueous medium. The self-assembly should arise from hydrophobic attractive interactions of PS segments forming a core, surrounded by a corona of protonated P2VP chains. In this context, the particle size distribution was investigated by dynamic light scattering (DLS) in the dilute aqueous regime (Figure S3). Two populations were observed, one corresponding to small aggregates (diameter < 100 nm) and another, more pronounced, which is attributed to large particles (100 nm < D < 350 nm). Graphite flakes were exfoliated in polymer solutions in acidic $H_2O$ (pH 2) at different concentrations (0.1 and 1.5 mg/mL). After isolation of the graphene/polymer (G/P) hybrids in the aqueous environment through the sonication/centrifugation protocol, transparent grey solutions were obtained (Figure 2).

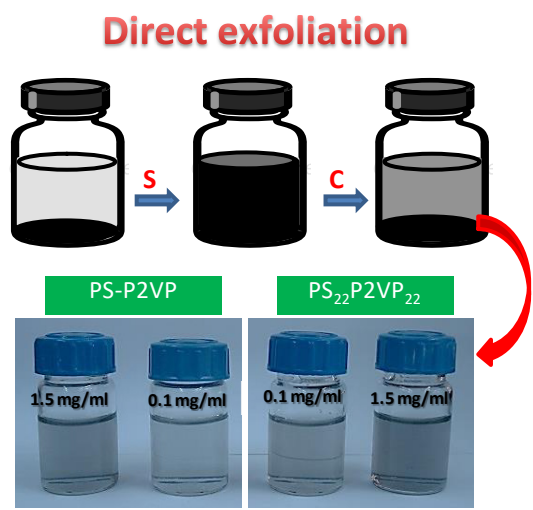

**Figure 2.** Direct exfoliation procedure of graphite in water of pH 2 (S: sonication, C: centrifugation) and digital photographs of G/P hybrid stable dispersions after the sonication/centrifugation cycle.

Estimation of G/P hybrid concentration in acidic water-based solutions was performed by adopting a weighting approach (Experimental Section). Both values of graphene/polymer concentration are presented in Table 2. The concentration of graphene exfoliated with the PS-P2VP polymers in $H_2O$ pH 2 was relatively low. This could be due to micellization of the polymer in the selective medium, with caging of PS chains into the hydrophobic core inhibiting the π-π interactions with the surface of graphene.

Table 2. G/P hybrid concentrations in various media.

| Sample | Polymer (mg/mL) | $C_{G/P}$/water pH 2 (mg/mL) | $C_{G/P}$/ethanol (mg/mL) | $C_{G/P}$/chloroform (mg/mL) |
|---|---|---|---|---|
| linear | 0.1 | 0.017 | 0.182 | 0.343 |
| linear | 0.5 | - | 0.216 | - |
| linear | 1.0 | - | 0.098 | - |
| linear | 1.5 | 0.034 | 0.058 | 0.377 |
| star | 0.1 | 0.021 | 0.114 | 0.178 |
| star | 0.5 | - | 0.131 | - |
| star | 1.0 | - | 0.068 | - |
| star | 1.5 | 0.042 | 0.056 | 0.312 |

In order to improve the concentration of graphene in solution, exfoliation in other selective media was attempted. The dispersability of carbon nanostructures was assessed through comparative optical observations of centrifuged graphene suspensions at the used media. Among the group of selective solvents used, ethanol was found to be the most efficient medium for obtaining suspensions with relatively high graphene concentration and four different polymer concentrations of 0.1, 0.5, 1.0 and 1.5 mg/m were examined.

The G/P hybrid concentrations are presented in Table 2. It is noted that, at high polymer concentrations (1.0 and 1.5 mg/mL) for both copolymers, graphene concentration decreased noticeably, possibly due to the formation of polymeric micelle aggregates, which are not able to penetrate within the galleries of graphene sheets, pre-exfoliated by the sonication process. The optimum polymer concentration for the graphene dispersability in ethanol was found to be 0.5 mg/mL, yielding for the linear copolymer 0.216 mg/mL G/P hybrids. Utilization of the linear polymer as graphene dispersant, yielded more concentrated graphene suspensions in all cases. This effect could be attributed to the multi-arm architecture of the star copolymer, which resembles a micelle configuration (Figure 1), inhibiting therefore the physical adsorption of the solvophobic domains onto graphene surface. Subsequently, dispersion in nonselective solvents was attempted, in order to avoid micellization phenomena. It was anticipated that the physical interactions between graphene surface and stabilizer would be maximized at lower polymer concentrations, compared with the case of selective solvents. Indeed, optical observations of the centrifuged samples showed that chloroform is the most efficient medium for obtaining concentrated graphene suspensions under similar treatment conditions. Suspensions containing both equal mass and equal molar quantities of polymer were prepared. Digital images of suspensions before and after centrifugation process are shown in Figure S4.

Estimation of G/P hybrid concentrations in chloroform-based solutions are presented in Table 2. Comparison of graphene dispersability in solutions of either linear or star polymer at the same mass concentration showed that the former type of stabilizer is more efficient particularly for the lowest polymer concentrations (0.1 mg/mL). The difference at higher polymer concentrations (1.5 mg/mL) was not so profound, indicating some possible saturation in graphene dispersability at relatively high stabilizer concentrations. This is strongly supported from the data extracted in the ethanol-based suspensions (see above). Comparison of equimolar polymer solutions as stabilizing agents for graphene exfoliation demonstrated slightly higher graphene dispersability in the linear polymer solution. Again a plausible explanation for the lower ability of the star copolymer as graphene stabilizer may involve the bulky character of the macromolecule resembling to micelle (incorporating 22 linear diblock copolymers), which could lead to less favoured physical interactions with the graphene basal plane.

*3.1.2 Film hydration*

As mentioned above, polymer micellization seems to hinder dispersability of graphene sheets in selective media. In order to avoid such phenomena in polymer solutions, alternative processing strategies should be adopted for enhancing the graphene dispersability in selective solvents. An alternative strategy towards the preparation of graphene suspensions in aqueous media involves the process of hydration of a graphene/polymer hybrid film. The precursor state of the latter was a stable chloroform suspension of graphene/polymer hybrid, in which the solvent was removed by evaporation. Due to the acidic environment of the added aqueous phase, the pyridine moieties are protonated and the graphene/copolymer hybrid nanostructures are extracted to the aqueous solution. A brief sonication results in further exfoliation of graphene multilayers, whereas the non-exfoliated material was discarded by centrifugation. Digital photos of centrifuged aqueous suspensions for both copolymers in two different concentrations are shown in Figure 3.

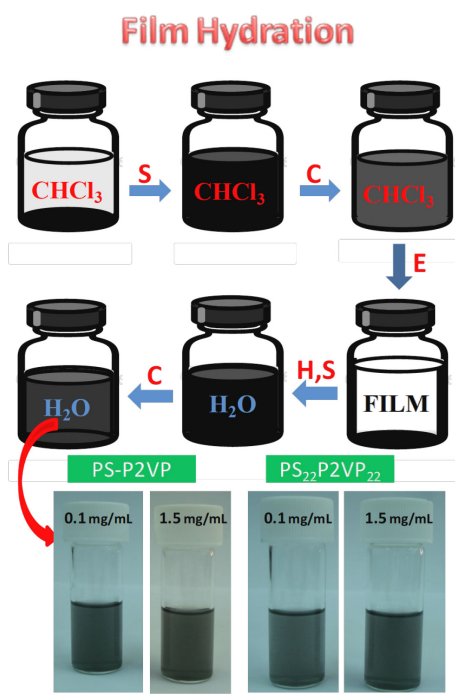

**Figure 3.** Film hydration exfoliation procedure of graphite in water of pH 2. (S: sonication, C: centrifugation, E: evaporation, H: hydration) and photographs of G/P hybrid stable dispersions after the last centrifugation step.

The concentrations of graphene sheets in the centrifuged aqueous suspensions were calculated gravimetrically and the values are shown in Table 3. As seen, relatively high G/P hybrid concentrations were obtained with this kind of extraction process. With respect to direct exfoliation of graphite in aqueous solutions of both copolymers, film hydration yields more than one order of magnitude higher concentrations of graphene suspensions. This should be attributed to the fact that unassociated macromolecules have been already adsorbed onto the graphene surfaces prior to film formation and hydration, since $CHCl_3$ is a good solvent for both blocks of the copolymer dispersing agent.

**Table 3.** G/P hybrid concentrations of the suspensions prepared by film hydration in $H_2O$ pH 2.

| Polymer | $C_P$ (mg/mL) | $C_{G/P}$ (mg/mL) |
|---|---|---|
| linear | 0.1 | 0.341 |
| linear | 1.5 | 0.370 |
| star | 0.1 | 0.180 |
| star | 1.5 | 0.316 |

Thanks to the reversibility of the protonation reaction of P2VP ($pK_a$ ca 5), the pH responsiveness of the composites in the acidic environment was evaluated by the addition of an appropriate volume of sodium hydroxide (NaOH) 1 M, so that the pH was set ~7. Mild agitation resulted in precipitation of the composites (floculates). Upon switching back the pH to acidic values (about 2) by addition of hydrochloric acid (HCl) 1M and tip sonication for 2 min, a homogeneous aqueous dispersion of the graphene/polymer hybrids was obtained (Figure S5).

*3.1.3 Two-phase graphene transfer*

An alternative processing strategy for producing stable graphene suspensions involves the shuttle transfer of graphene sheets between immiscible media. We studied the effect of polymer architecture (linear vs star) with regards to the dispersability and exfoliation efficiency of graphene after shuttle transfer between chloroform and acidic water. Initially, we evaluated the influence of acidity of the receiving aqueous solution to the efficiency of shuttle process itself. On the top of centrifuged graphene/polymer suspensions in $CHCl_3$, an equal volume of aqueous medium of different pH values (1, 2, 3, 4 and 5.5) was added in the receiving compartment. After moderate agitation (100 rpm) for 72 h, the aqueous phase was separated (Figure S6) and was assessed by optical means. As observed, only the aqueous environment at pH values of 1 and 2, was able to act as an efficient receiving medium for the graphene composites, due to adequately ionization of the P2VP segments of copolymer. In addition, this shuttle transfer process has resulted in an increase in the pH of the final aqueous dispersion from pH 1 to 1.74 and from pH 2 to 2.48, owing to the protonation reaction of the P2VP blocks of the adsorbed copolymers onto the graphene surface.

Shuttle transfer from organic to acidic aqueous solutions was roughly completed after 3 days for most of the samples. Optical observations of the receiving aqueous suspensions (Figure 4) showed that, under the specific conditions, graphene material does not transfer at comparable rates from the organic solution of star copolymer at the lowest concentration (0.1 mg/mL). In fact, a very low fraction of graphene sheets dispersed in chloroform was transferred in the upper phase. The enhanced efficiency of the less bulky linear block copolymer for shuttling the graphene nanostructures could be interpreted by the presence of a higher number of polymeric chains that can be potentially adsorbed onto a specific graphene surface.

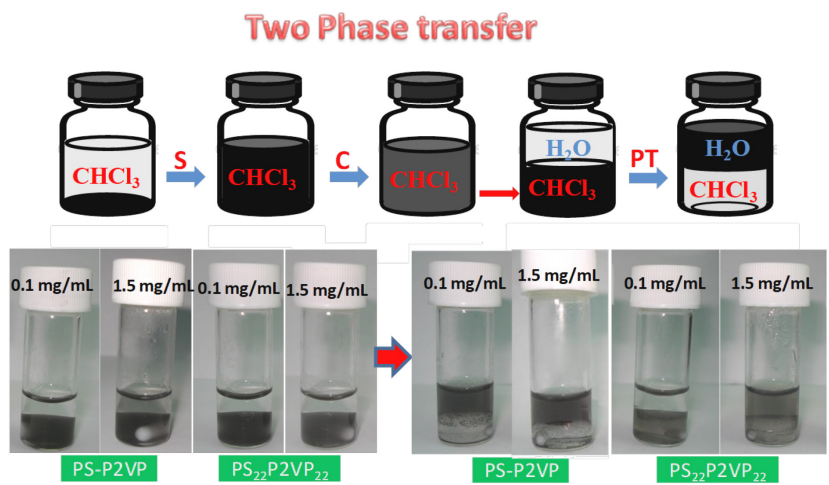

**Figure 4.** Two step procedure of graphite exfoliation in water: direct exfoliation in CHCl$_3$ followed by phase transfer in water of pH 2 (S: sonication, C: centrifugation, PT: phase transfer) and photographs of G/P hybrid phase transfer for various samples.

Absolute values of graphene hybrid concentrations in the acidic aqueous suspensions were calculated by a combination of gravitational approach and by recording the UV-Vis spectra (Figure S2), along with the estimation of absorption coefficient values at a specific wavelength of 660 nm (Table S2). The values of graphene concentrations after the shuttling process to the acidic aqueous suspensions are given in Table 4. The data clearly show that by comparing equimolar polymer solutions, 0.1 mg/ml (L) vs 1.5 mg/ml (S), the linear copolymer is more efficient in transferring graphene at the aqueous phase. It seems that the ability of linear copolymer solution to transfer graphene saturates at the lower polymer concentration (0.1 mg/mL), although different samples with variable block copolymer concentrations should be tested in order to get an optimum condition. On the contrary, in the case of star copolymer, great enhancement of the graphene transfer ability was observed at a 1.5 mg/mL polymer concentration. If we assume that the star resembles a micelle constituted of linear copolymers with $N_{agg}$ equal to the number of each type of arms (n=22), then this could explain why the star of the same low concentration (0.1 mg/ml) was less efficient in exfoliating a reasonable quantity of graphite as compared to direct exfoliation (Figure 2).

Table 4. G/P hybrid concentration in the receiving aqueous compartments prepared by two-phase shuttle transfer.

| polymer | $C_P$ (mg/mL) | $C_{G/P}$ (mg/mL) |
|---|---|---|
| linear | 0.1 | 0.287 |
| linear | 1.5 | 0.284 |
| star | 0.1 | 0.014 |
| star | 1.5 | 0.219 |

In order to monitor the kinetics of graphene shuttle transfer from chloroform to acidic water, UV-Vis spectroscopy was used. Specific volumes of either chloroform or aqueous aliquots were sampled out at various time intervals during the 72 h period of shuttle process. We observed that graphene/polymer hybrids were transferred to the aqueous phase at different rates, depending on polymer concentration and architecture (Figure 5).

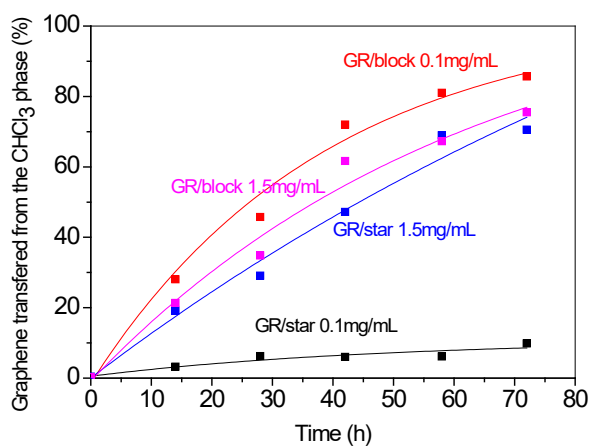

**Figure 5.** Kinetics of phase transfer from $CHCl_3$ to acidified water. The Lines guide the eyes.

In the case of linear diblock copolymer, the rate of graphene transfer was higher when the starting polymer concentration was 0.1 mg/mL. The corresponding rates for the star copolymer were lower and similar to those of the linear counterpart, with the 1.5 mg/mL sample being transferred at a clearly more efficient manner. It can be observed that, while the graphene hybrid is not able to transfer appreciably (corresponding to $1.6 \times 10^{14}$ star molecules per volume) from a 0.1 mg/ml star copolymer solution in $CHCl_3$, an increase of about one order of magnitude in polymer concentration (1.5 mg/ml, corresponding to $2.4 \times 10^{15}$ chains per volume), enabled phase transfer. On the other hand, at similar molar concentration for the linear copolymer ($2.4 \times 10^{15}$ molecules per volume), the phase transfer proceeds at a faster rate relatively to the star counterpart, while at enhanced mass concentration of 1.5 mg/ml ($3.7 \times 10^{16}$ molecules per volume) corresponding to 15-fold increase, no significant improvement in the rate of phase transfer was observed.

In a subsequent step, we studied the reversibility of the shuttle transfer process. The pH of the aqueous media was increased to 6.8, upon addition of appropriate volume of NaOH 0.1 M (~20 μL). Further agitation resulted in a back transfer of the hybrids to the $CHCl_3$ phase, since the P2VP segments were completely deprotonated, transformed to hydrophobic (Figure 6). In order to test to what extent the composites may be transferred back and forth, the pH of the upper aqueous phase was readjusted to pH 2 by addition of appropriate volumes of HCl 0.1 M (~38 μL). After moderate agitation, the hybrids were transferred again to the acidic aqueous phase.

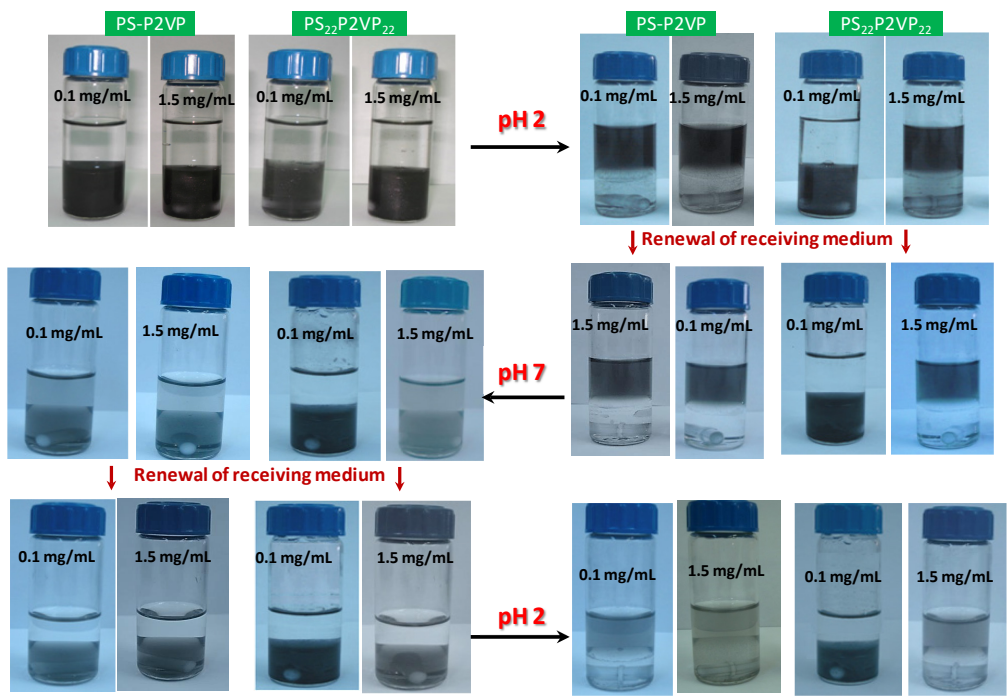

**Figure 6.** Digital photographs showing the G/P hybrid shuttle between $CHCl_3$-$H_2O$ phases triggered by pH. In the first step (first line) the graphene sheets were transferred to acidified water (pH 2). In the second step (second line), back transfer to $CHCl_3$ occurred by switching the pH of water phase to 7. Finally the graphene sheets were again transferred to water pH 2 (third line). In each step the receiving medium was renewed.

In order to assess the transfer efficiency of the G/P hybrids in other media by the two-phase transfer method, an ionic liquid was investigated as a potential receiving compartment. In the first setup, the carbon nanostructures suspended in $H_2O$ of pH 2 through the film hydration protocol, were further transferred in ionic liquid (IL) 1-butyl-3-methylimidazolium hexafluorophosphate [BMIM][PF6] by gentle stirring (Figure 7). The amount of graphene transferred from the aqueous phase to the IL was monitored by sampling out aliquots of both aqueous and ionic liquid suspensions and by recording their UV-Vis spectra, as mentioned previously. The concentration of G/P hybrid transferred to the ionic liquid phase is presented in Table 5.

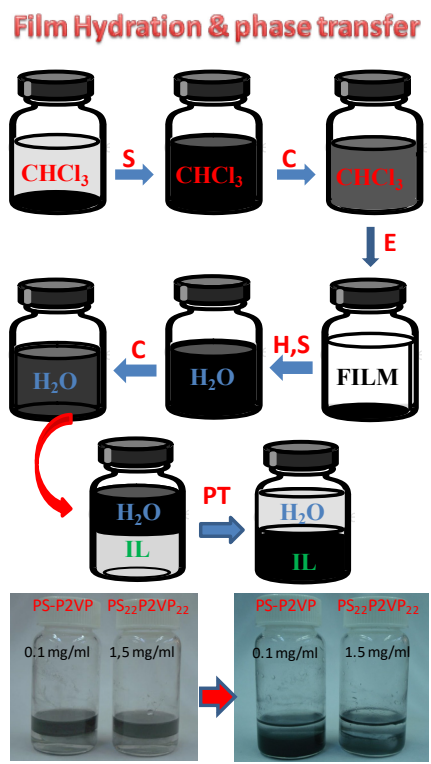

**Figure 7.** Multistep procedure of graphite exfoliation in IL (setup No1) and digital photographs of G/P hybrids transferred from H$_2$O (pH 2) to IL in the last step (S: sonication, C: centrifugation, E: evaporation, H: hydration, PT: phase transfer)

**Table 5.** G/P hybrid concentration transferred from H$_2$O pH 2 to IL determined by UV.

| polymer | C$_P$ (mg/mL) | C$_{G/P}$ (mg/mL) (setup No1) | C$_{G/P}$ (mg/mL) (setup No2) |
|---|---|---|---|
| linear | 0.1 | 0.299 (87%) | 0.157 (55%) |
| star | 1.5 | 0.243 (78%) | 0.088 (40%) |

*In parenthesis (transferring percentage from the water phase)*

In the second setup (Figure 8), the graphene/polymer hybrid nanostructures were firstly transferred from CHCl$_3$ to H$_2$O pH 2 and subsequently to the ionic liquid [BMIM][PF$_6$]. The concentration of graphene hybrids in the receiving IL phase is shown in Table 5. By comparing the graphene concentrations in the final IL phase by both approaches, it was clearly seen that the former setup (film hydration and phase transfer) the percentage of the graphene hybrids transferred from the aqueous phase were appreciably higher yielding a high exfoliation yield of carbon nanostructures in [BMIM][PF6].

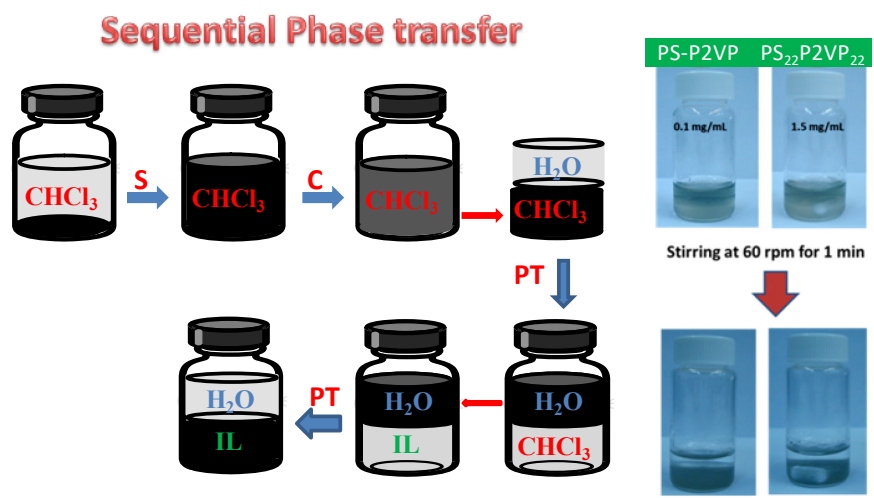

**Figure 8.** Multistep sequential phase transfer of G/P hybrids from $CHCl_3$ to acidic water and then to IL (setup No2) and digital photographs of G/P hybrids transferred from $H_2O$ (pH 2) to IL (S: sonication, C: centrifugation, PT: phase transfer).

Note that direct exfoliation of graphene in IL in the presence of PS/P2VP copolymers did not yield stable graphene suspensions likely because the P2VP blocks were not protonated.

*3.1.4. True exfoliated graphene concentration, polymer adsorption and yield*

The $C_{G/P}$ (mg/mL) determined by UV-vis concerns the concentration of the hybrids, that is, graphene including the adsorbed polymer. Thus, the determined absorption coefficients of the hybrids $α_{G/P}$ are apparent (SI) and different in every experiment, as well as, from that of neat graphene ($α_G$=3620 $Lg^{-1}$ $m^{-1}$).[19,20,35,36] Considering that the decrease of the transmiting light through the suspensions is due to the suspended graphene sheets (the effect of the adsorbed macromolecules should be negligible), a number of important factors concerning a comprehensive characterization of the obtained graphene/polymer hybrids can be evaluated. The percentage of the adsorbed polymer, *p.a.(%)*, onto the exfoliated graphene sheets can be calculated by the formula:

$$p.a.(\%) = (1 - \frac{c_G}{c_{G/P}}) \times 100$$

where

$$\frac{c_G}{c_{G/P}} = \frac{A/la_G}{A/la_{G/P}} = \frac{a_{G/P}}{a_G}$$

Therefore, the neat graphene concentration can be determined by the equation $c_G = c_{G/P}(\alpha_{G/P}/\alpha_G)$ and thus can be compared with other liquid exfoliation procedures. Other parameters that could be extracted from the above calculations is the exfoliation yield, that is $(c_G/c_{GF}) \times 100$, (where $c_{GF}$ is the initial graphite concentration used), as well as, the mass of exfoliated graphene to the mass of the adsorbed polymer ratio G/P $[c_G/c_{G/P} \times p.a.)]$. All these parameters, which are deemed important for evaluating potential technological applications, are presented in Table 6.

**Table 6.** Quantitative characterization of Graphene/Polymer hybrids

| polymer | medium (procedure)[a] | $C_P$ (mg/mL) | GF/P[b] ratio | $C_{G/P}$ (mg/mL) | $C_G$ (mg/mL) | Yield[c] (%) | p.a (%) | G/P ratio |
|---|---|---|---|---|---|---|---|---|
| linear | CHCl$_3$ (DE) | 0.1 | 20 | 0.341 | 0.194 | 9.7 | 43.2 | 1.3 |
| linear | CHCl$_3$ (DE) | 1.5 | 4/3 | 0.370 | 0.275 | 13.7 | 25.7 | 2.9 |
| linear | H$_2$O (PT) | 0.1 | 20 | 0.287 | 0.105 | 5.2 | 63.4 | 0.6 |
| linear | H$_2$O (PT) | 1.5 | 4/3 | 0.284 | 0.140 | 7.0 | 50.7 | 1.0 |
| star | CHCl$_3$ (DE) | 0.1 | 20 | 0.180 | 0.123 | 6.2 | 31.6 | 2.2 |
| star | CHCl$_3$ (DE) | 1.5 | 4/3 | 0.316 | 0.174 | 8.7 | 45.0 | 1.2 |
| star | H$_2$O (PT) | 0.1 | 20 | 0.014 | 0.006 | 0.3 | 57.1 | 0.8 |
| star | H$_2$O (PT) | 1.5 | 4/3 | 0.219 | 0.128 | 6.4 | 41.6 | 1.4 |

[a] (DE) direct exfoliation, (PT) phase transfer
[b] initial graphite to polymer ratio
[c] exfoliation yield with respect to the initial $c_{GF}=2$ mg/ml

## 3.2 Qualitative evaluation of exfoliated graphene

*3.2.1 TEM-SEM*

After evaluating the graphene dispersability at various media by the direct or indirect strategies (weighting and/or optical approach) reported above, we then focused our efforts on the assessment of exfoliation efficiency. TEM imaging may give useful information about the exfoliation efficiency of graphene/polymer hybrids, due to the electron transparency of the polymer component. It was shown that direct exfoliation of graphene in selective media, such as acidic water (pH 2) and ethanol was an inefficient route for obtaining adequately exfoliated material as thick non-transparent objects were found to be present in the solution (Figure S7a). More transparent platelets were recorded in the case for which ethanol was used for direct exfoliation (Figure S7b). It is noted here that there is no profound difference in the morphology of deposited graphenes, when comparing hybrids either based on the linear (L) or star (S) copolymer, at all used polymer concentrations.

In the case of ethanol-based suspensions, the influence of the starting mass ratio of graphite to polymer on the morphology and lateral size of exfoliated graphene nanostructures was investigated. As shown in Table S3, graphene suspensions with four different component mass ratios were prepared. TEM and SEM imaging clearly showed that the lowest the "graphite to polymer" weight ratio, the larger are the graphene sheets isolated by the casting process. Especially for the suspension with graphite to copolymer ratio value of two (2), large graphene nanostructures were remained in suspension, even after centrifugation up to 2500 rpm (Figure S8). The stability of graphene sheets after centrifugation at such rates demonstrated that the suspended platelets are rather few-layer structures of lateral size between 2 and 30 μm. On the contrary, as the component ratio increased to 20, the mean size of deposited graphenes was decreased towards the sub-micron range.

Direct exfoliation of graphene in chloroform solutions of both copolymers yielded few-layer carbon-based nanostructures regardless of stabilizer architecture (Figure S7c,d). On the contrary, the material which was prepared through the film hydration process was mostly multi-layer structures, similar to those suspended in acidic water after direct exfoliation (Figure 9 a,b). The lack of adequate exfoliation could be ascribed to the fact that the starting graphene/polymer hybrid material was in the form of dried film, which rendered less efficient the exfoliation of such multi-layered aggregates by brief sonication. Concerning the effect of polymer architecture on the exfoliation efficiency of graphene sheets after a two-phase transfer from chloroform to acidic water, it was observed that, on average, graphene/star copolymer hybrids were more transparent than those obtained by using linear copolymer as a stabilizer (Figure 9, images c, d and Figures S9 and S10).

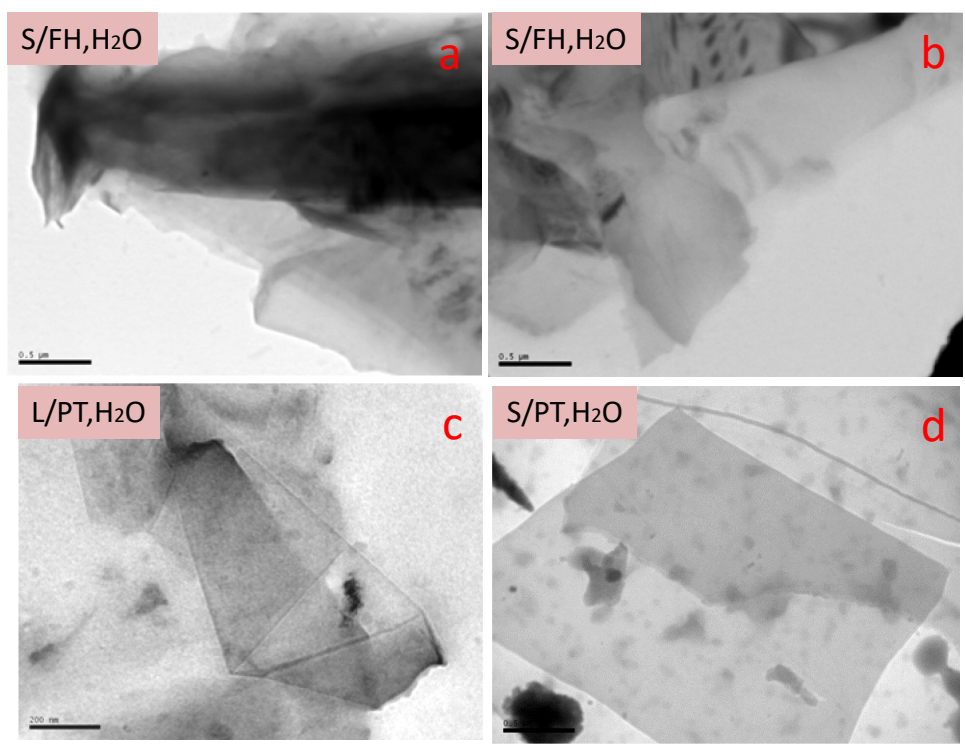

**Figure 9.** Representative TEM images from the film hydration (FH) method using linear (L), (a) or star (S) copolymer, (b) and from the phase transfer (PT) method from $CHCl_3$ to acidic water using linear (c) or star copolymer (d)

It is noted here that there is no obvious effect of starting polymer concentration. Thus, it seems that although the graphene/linear copolymer system results in more concentrated suspensions in the receiving aqueous phase, the exfoliation quality is on average less efficient, when comparing with the star counterpart. The TEM images suggest that in the latter case monolayer, two-layer and few-layer graphenes are produced by the phase transfer method.

*3.2.2 Raman analysis*

In order to corroborate the exfoliation efficiency of suspended graphenes in various systems by TEM imaging, we performed a mapping of the Raman spectra of the carbon nanostructures deposited onto $SiO_2$/Si wafers. Raman is a very useful tool for characterization of graphitic materials, namely their graphitization level, relative population of defect sites, as well as, the identification of graphene layer for a given specimen.[37-39] The latter parameter may be evaluated both by the shape of the so-called 2D peak at about 2700 $cm^{-1}$, as well as, the intensity ratio of G (at 1600 $cm^{-1}$) to 2D peak. In Figure 10A, the Raman spectrum of pristine graphite is shown, where the characteristics graphene peaks are shown. Due to the large size of crystallites and the multi-layer character, there is no appreciable D peak at about 1350 $cm^{-1}$, which emanates from defect sites or high concentration of edges.

Concerning the case of direct exfoliation of graphene in both selective and nonselective media, Raman mapping was carried out in graphene nanostructures dispersed in either chloroform, ethanol or acidic water. In the chlorinated medium, the layer number of deposited graphenes was ranged between 3 and 8. A representative

Raman spectrum of a trilayer structure is shown in Figure 10B. It is worth noting that no significant differences were observed for the average layer number of graphenes exfoliated either by the block or the star copolymer at both polymer concentrations. The integration ratio of 2D and G peaks (2D/G) was found to range between 1.2 and 2.6, depending on the layer number, whereas the corresponding ratio D/G was between 0.12 and 0.16. In ethanol-based suspensions, the average distribution of layer number was slightly shifted towards higher values (4-8) (spectrum not shown). Direct exfoliation in acidic water has not provided any efficient exfoliation, as is evident by TEM imaging (Figure S7a). Raman mapping demonstrated that multi-layer graphenes of layer number > 10 were deposited onto the Si/SiO$_2$ wafers (spectrum not shown).

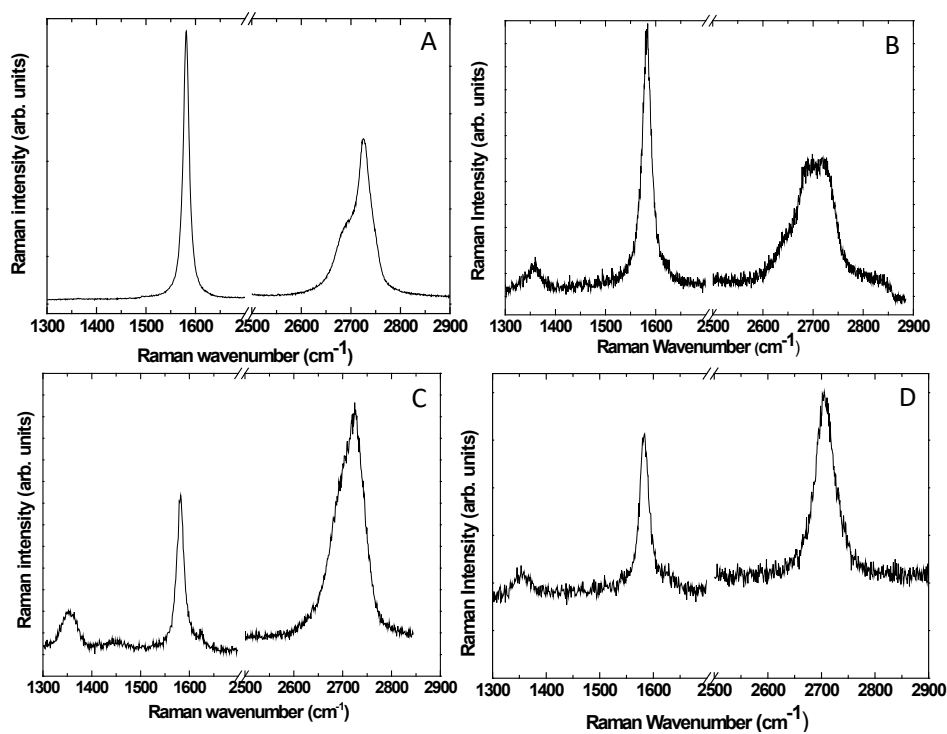

**Figure 10.** Representative Raman spectra (514 nm) of deposited graphene flakes from: (A) pristine graphite; (B) direct exfoliation in CHCl$_3$; (C) two-phase transfer of linear block copolymer and (D) of star copolymer-based hybrid from CHCl$_3$ to acidic water.

Similar results were obtained for the carbon nanostructures derived by the "film hydration" protocol, where again multi-layer graphenes were observed as is evident by their Raman spectra which resembled that of pristine graphite (Figure 10A). The Raman data were also strongly corroborated by the TEM images (see Figure 9a,b), in which non-transparent graphene structures were observed.

In the aforementioned strategies of both direct exfoliation and film hydration, there were no noticeable differences, when varying either the copolymer architecture or the concentration. On the contrary, optical characterization of graphene sheets derived from a two-phase transfer process gave rise to different exfoliation efficiency, depending on the architecture of stabilizer. After pre-exfoliation of graphite in the organic medium and gentle stirring of the biphasic systems for about 72 h, the graphene nanostructures were transferred to the aqueous phase. Mapping of the deposited graphenes derived from each aqueous sample showed that graphene hybrids stabilized with the linear copolymer were mostly few-layer sheets (layer number 3-8). A representative Raman spectrum can be seen in Figure 10C. On the contrary, in the material stabilized by star copolymer chains, an appreciable fraction of deposited graphenes were monolayers. This implies that few-layer graphene sheets in the organic phase were further exfoliated during the shuttle transfer process. Statistical analysis demonstrated that about 15% of graphene/star copolymer hybrids were monolayers (Figure 10D), whereas the remaining nanostructures varied between 2-4 layered graphenes.

## 3.3 Fabrication of graphene/copolymer/PVA composite films

Since the two-step process, involving graphite pre-exfoliation in $CHCl_3$ followed by phase transfer in aqueous media, was the best graphene dispersion method from both quantitative and qualitative point of view, it was interesting to assess the potential of this exfoliation method to the preparation of functional polymer nanocomposites for mechanical reinforcement. For this purpose we prepared PVA-graphene membranes with 0.1 wt% graphene content (see details in SI). PVA was chosen as the matrix component, due to its compatibility with the P2VP arms of the star copolymer $PS_{22}P2VP_{22}$ and also for comparison purposes with previous studies.[34,40,41] This star copolymer played not only the role of exfoliation agent of pristine graphite in solution but it also acts as a compatibilizer between the P2VP arms and the hydroxyl groups of PVA, through hydrogen bond formation. Thus, it is anticipated that enhanced adhesion between the filler and the matrix may be built within the composite.

The results of the DMA experiments on the graphene/PVA are given in Figure 11. The incorporation of only 0.1 wt% of graphene in the PVA matrix resulted in an increase by 145% (at $-50^0$ C) of the E' value from 2.7 GPa (neat PVA) to 6.6 GPa. To the best of our knowledge, this is the highest storage modulus obtained for a PVA/graphene nanocomposite containing just 0.1 wt% graphene. At room temperature (25 $^o$C), the storage modulus increases from ~2.2 GPa to ~4.2 GPa at a testing frequency of 10 Hz. This again is a significant increase for weight fractions of 0.1% and far surpasses the observed increase of the static (i.e. zero frequency) tensile Young's modulus for a similar graphene/ PVA composite.[34] According to composite mechanics, the significant improvement over previous attempts observed in this work can mainly be attributed to the size and orientation of the inclusions i.e. the graphene flakes. In order to estimate the average length of the graphene flakes, we examined the fracture surface of the nanocomposite by SEM.

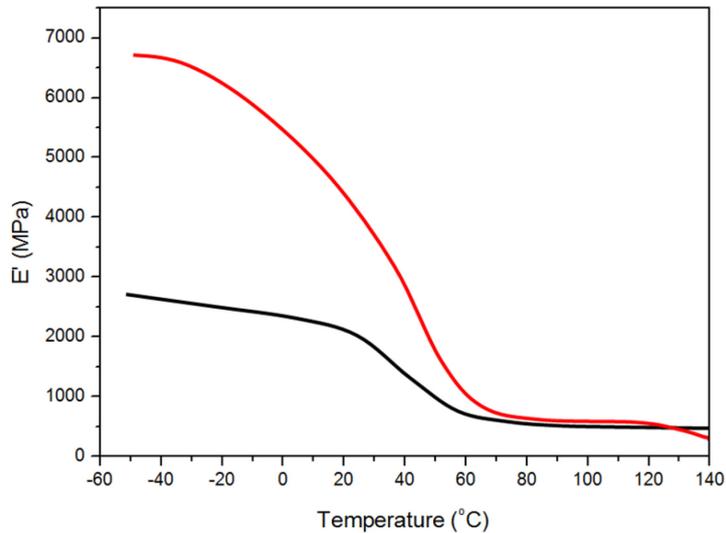

**Figure 11:** Temperature dependence of storage modulus for the neat PVA (black line) and the nanocomposite film (red line).

As seen in Figure 12 (and in Fig. S10), the edges of each graphene layer are clearly visible and the graphene flakes appear transparent (monolayers) and be randomly dispersed into the polymer matrix. Moreover, the mean size of the graphene flakes is of the order of several μm, as estimated from the SEM micrographs in good agreement with TEM results after phase transfer (Figure 9d). This is indeed an important result since it appears that the preparation conditions (liquid exfoliation, sonication, phase transfer) described previously, did not affect adversely the graphene size leading to nanocomposites containing non-agglomerated and relatively large graphene flakes. Based on these results, we believe that the large size of flakes ensures efficient stress transfer from polymer to the graphene flakes during loading. Indeed, recent work [42, 43] has shown that for engineering matrices, a transfer length of at least 2 μm is required for efficient stress transfer.

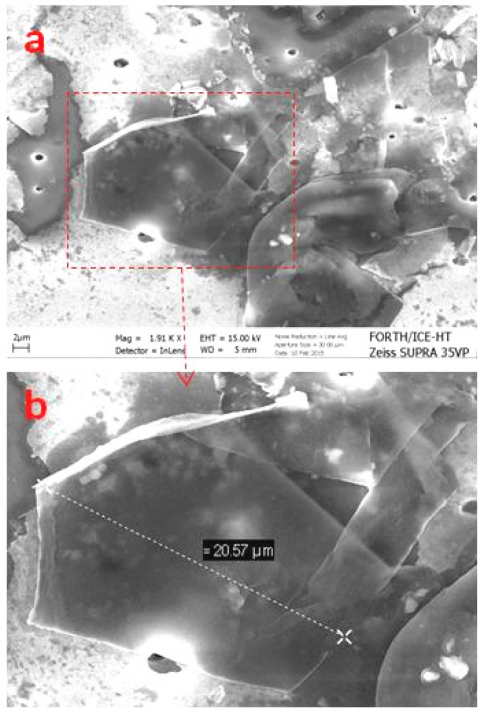

**Figure 12:** SEM micrograph (a) obtained from the cryofracture surface of PVA/graphene nanocomposite film and blow up of the marked (dashed line) area (b).

Hence, in our case, the effect of transfer length is minimal and therefore the values of modulus obtained reflect mainly the volume fraction of the nanocomposite and not interfacial issues such as the transfer length. It is also evident that flakes smaller than 2 μm cannot provide full reinforcement and this is why in previously reported cases [34] only a marginal improvement over the matrix modulus can be achieved. Secondly, due to the approximately rectangular shape of the flakes and their large size, there is a tendency of the flakes to orient themselves parallel to each other and to the applied stress. This again ensures efficient transfer of the applied load to the graphene flakes and explains further our results. Current work is under way to verify fully the above assumptions for a whole range of graphene volume fractions.

Finally, thermal characterization of the prepared thin films was determined by Differential Scanning Calorimetry (DSC). In comparison to neat PVA, the Tg of the nanocomposite with 0.1 wt % filler loading slightly increased about 1.5 $^{o}$C (Figure S11) which is attributed to the reduced mobility of polymer chains due to the effective attachment of PVA to the nanosheets of graphene.[44] More importantly, the $T_m$ of the nanocomposite was found to decrease significantly (about 27 $^{o}$C) implying that incorporation of graphene sheets into PVA matrix influences remarkable the crystallization behavior of PVA, leading to the formation of polymer crystals of smaller size and perfection.

## 4. Conclusions

In this work, exfoliation and stabilization of few-layer graphene nanosheets has been attempted by using linear PS-b-P2VP and heteroarm star PS$_{22}$P2VP$_{22}$ asymmetric (ca 80% P2VP) ionizable block copolymers. For this purpose three different processing strategies, towards the preparation of efficiently exfoliated graphene aqueous dispersions, were attempted and compared with. The processes involved short-time tip sonication and direct exfoliation, film hydration or phase transfer. The choice of the second P2VP block was critical since it can be transformed from lipophilic to hydrophilic by reversible protonation while it is soluble in organic media in the deprotonated form (Figure 13). Direct exfoliation of graphene to acidic aqueous media was poor due to the antagonistic micellization process occurring to the amphiphilic block copolymer dispersing agents. On the contrary, phase transfer from organic (common good solvent for the different blocks of the copolymer) to acidic aqueous phase was shown to be the most efficient method for acquiring adequately exfoliated graphene sheets in solution. The alternative film hydration method yielded high concentration graphene suspensions, yet, with low exfoliation efficiency (i.e. few layer graphenes).

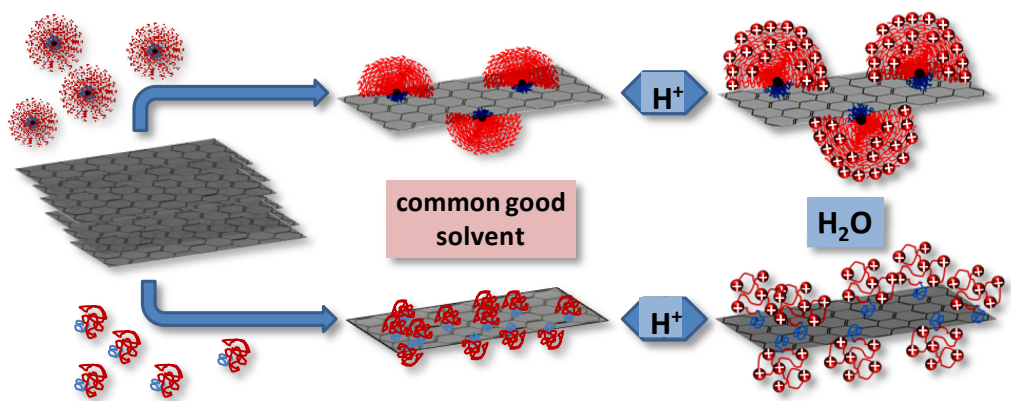

**Figure 13**. Schematic representation of exfoliation of graphite using linear (PS-P2VP) and star-shaped (PS$_n$-P2VP$_n$) block copolymers as dispersing agents in organic solvent and transferring to water by transforming the lipophilic P2VP block/arms to hydrophilic by reversible protonation.

Polymer architecture seemed to play a role in the shuttle transfer process. The linear copolymer was more efficient in obtaining high concentration graphene aqueous suspensions, yet, the heteroarm star copolymer seemed to be a better exfoliation agent, leading to the production of aqueous suspensions which are highly enriched in monolayer graphenes. More importantly, the short-time sonication ensured that graphenes of μm dimensions were obtained which, in turn, is a prerequisite for efficient stress transfer in polymer/graphene nanocomposites. The phase transfer process was found to be reversible between organic and aqueous media thanks to the protonation/deprotonation equilibrium of P2VP controlled by pH. Thus a reasonable yield of "smart" graphene/polymer hybrids were obtained that could be employed for water purification since they could absorb pollutants and can be then easily removed from water by increasing pH.

Moreover, sequential phase transfer of graphene sheets from acidic aqueous medium to ionic liquid was proved to be a rapid process, leading to nearly quantitative mass transport. Therefore, with the same polymer as exfoliating and dispersing agent it is possible to produce graphene/polymer hybrid suspensions in three different media , namely (low boiling point) organic solvents, water and ionic liquids.

Concerning the development of functional polymer composites, the graphene/copolymer hybrid, suspended in aqueous medium by the star copolymer, was compounded with PVA matrix, giving rise to enhanced mechanical reinforcement of the polymer, by using only 0.1 wt% of filler material. The obtained significant increase of storage modulus up to 6.6 GPa (245% higher than that of neat PVA) is attributed to the large size and rectangular shape of well- dispersed inclusions (mono and few layer graphens) that resulted in efficient stress transfer under load and also uniform orientation along the loading direction.

**Acknowledgments**

This research has been co-financed by the European Union (European Social Fund (ESF)) and Greek national funds through the Operational Program "Education and Lifelong Learning" of the National Strategic Reference Framework (NSRF)-Research Funding Program: Thales: Investing in knowledge society through the European Social Fund. Finally one of us (CG) wishes to acknowledge the financial support of the Graphene FET Flagship (''Graphene-Based Revolutions in ICT And Beyond''- Grant agreement no: 604391) and of the European Research Council (ERC Advanced Grant 2013) via project no. 321124, "Tailor Graphene".